\documentclass{amsart}

\theoremstyle{definition}

\theoremstyle{remark}

\def\Section{\section}
\def\SubSection{\subsection}
\def\qq{\quad}
\def\q{\quad}
\def\predel{\lim}
\def\rsp{\bf R}
\def\e{{\it e}}
\def\i{{\it i}}
\def\d{{\it d}}
\def\ddd{{\it d}}
\def\no{{\it N}}
\def\ref{\cite}
\begin{document}

Jour. of Inverse and ill-posed problems,
Vol. {\bf 7}, N 6, (1999), p. 561-571.
\vspace{0.5cm}

\title[Inverse scattering problem]{An approximate method for solving the inverse \\ 
scattering problem with fixed-energy data}
\author[A.G.~Ramm and W.~Scheid]{A.\,G.~Ramm\footnote{
    Mathematics Department, Kansas State University, Manhattan, KS66506-2602, USA.\\
    E-mail:~ramm@math.ksu.edu} and
W.~Scheid\footnote{
    Institut f\"ur Theoretische Physik der Justus-Liebig-Universit\"at Giessen,
    Heinrich-Buff-Ring 16, D 35392, Giessen, Germany.
    E-mail:~werner.scheid@theo.physik.uni-giessen.de\\[2mm]
    The work was supported by DAAD.}}
\begin{abstract}
Assume that the potential $q(r)$, $r>0$, is known for $r\ge a>0$, and
the phase shifts $\delta_l(k)$ are known at a fixed energy, that is at a
fixed $k$, for $l=0,1,2,\ldots\,$. The inverse scattering problem is: find
$q(r)$ on the interval 0$\le r\le a$, given the above data.
A very simple approximate numerical method is proposed for solving
this inverse problem. The method consists in reduction of this problem
to a moment problem for~$q(r)$ on the interval $r\in [0,a]$. This moment
problem can be solved numerically.
\end{abstract}

\maketitle

\Section{Introduction}
\setcounter{equation}{0}
\renewcommand{\theequation}{1.\arabic{equation}}
\vspace*{-0.5pt}   

Finding a potential $q(r)$, $r=|x|$, from the phase shifts
$\delta_l(k)$ for angular quantum numbers $l=0,1,2,\ldots\,$, known at a
fixed wave number $k>0$, that is, at a fixed energy, is of interest in
many areas of physics and engineering. A~parameter-fitting procedure for
solving this problem was proposed in the early sixties by R.\,G.~Newton
and discussed in \ref{r3,r2}. This procedure has principal drawbacks
which have been discussed in detail in the paper \ref{r1}.

In [1] AGR pointed out for the first time
that the Newton\,--\,Sabatier method widely used
by the physicists to invert fixed-energy phase shifts
for the potential is not really an inversion method but a
parameter-fitting procedure; that this procedure
is not always applicable to the data; and that
it is not proved that this procedure leads to the original potential,
which produced the original phase shifts.
Moreover, the Newton\,--\,Sabatier procedure, if it is applicable,
can only produce a potential which is analytic
in a neighborhood of $(0,+\infty)$, with a possible pole at $r=0$.
A general numerical method is given in~[1]
to construct piecewise-constant compactly supported
spherically-symmetric potentials which are quite different
but have nearly the same sets of phase shifts
at a fixed energy.

If $k> 0$ is large, then one can use a stable numerical inversion
of the fixed-energy
scattering data in the Born approximation \ref[Section~5.4]{r4}. 
Error estimates
for the Born inversion are obtained in \ref{r4}.

An exact (mathematically rigorous) inversion method for solving a 3D
inverse scattering problem with fixed-energy noisy data is developed in
\ref{r5} where error estimates were derived and the stability of the solution
with respect to small perturbations of the data was estimated.
This method, although rigorous, is not very simple.

The aim of this paper is to propose a novel, approximate, quite simple in
principle, method for inverting the data $\{\delta_l\}$, $l=0,1,2,\ldots\,$,
given at a fixed $k> 0$, for the potential $q(r)$. In most physical
problems one may assume that~$q(r)$ is known for $r\ge a$ (near infinity), and one
wants to find $q(r)$ on the interval $[0,a]$, where $a> 0$ is some known radius.

Our basic idea is quite simple: since $\delta_l$ and $q(r)$ for $r\ge a$
are known, one can easily compute the physical wave function
$\psi_l(r)$ for $r\ge a$, and so the data
\begin{equation}
\{\psi_l(a), \psi'_l(a)\},\qq  l=0,1,2,\ldots
\end{equation}
can be obtained in a stable and numerically efficient way (described in  Section~2 below)
from the original data $\{\delta_l\}$, $\{q(r)$ for $r\ge a\}$.

Now we want to find $q(r)$ on the interval $r\in[0,a]$ from the data (1.1).
The function $\psi_l(r)$ on the interval $[0,\infty)$ solves the integral equation
\begin{equation}
\psi_l(r) = \psi_l^{(0)}(r)-\int^a_0 g_l(r,\rho)q(\rho)\psi_l(\rho)d\rho,\q 0\le r\le a
\end{equation}
where $g_l$ is defined in (A.12) and $\psi_l^{(0)}(r)$ is a known
function which is written explicitly in formula (2.7) below. Let
\begin{equation}
-[\psi_l(a)-\psi_l^{(0)}(a)]:=b_l,\quad
-[\psi_l^{\prime}(a)-\psi_l^{(0)\prime}(a)]:=\beta_l, \quad l=0,1,\ldots\,.
\end{equation}
The numbers $b_l$ and $\beta_l$ are known. Taking $r=a$ in (1.2)
and approximating $\psi_l$ by $\psi_l^{(0)}$ under the sign of the integral
one gets:
\begin{equation}
\int^a_0 q(\rho)f_l(\rho)d\rho = b_l,\quad l=0,1,2,3,\ldots
\end{equation}
where
\begin{equation}
f_l(\rho): = g_l(a,\rho)\psi^{(0)}_l(\rho).\nonumber
\end{equation}
Differentiate (1.2) with respect to $r$, set $r=a$, and again
replace $\psi_l$ by $\psi_l^{(0)}$ under the sign of the integral. The result is:
\begin{equation}
\int^a_0 q(\rho)s_l(\rho)d\rho = \beta_l, \quad l=0,1,2,3,\ldots
\end{equation}
where
\begin{equation}
s_l(\rho): =\predel{\frac{\partial g_l(r,\rho)}{\partial r}}_{r=a}\psi_{l}^{(0)}(\rho).
\end{equation}
We have replaced  $\psi_l(\rho)$ in (1.2) under the sign of the integral
by $\psi^{(0)}_l(\rho)$. Such an approximation is to some extent similar
to the Born approximation and can be justified if $q(\rho)$ is small or
$l$ is large or $a$ is small.
Note that this approximation is also different from the Born
approximation since $\psi_l^{(0)}$ is defined by formula~(2.7) and
incorporates the information about $q(r)$ on the interval $r\ge a$.

For potentials which are not small and for $l$ not too large such an
approximation cannot be justified theoretically, but may still lead to
acceptable numerical results since the error of this approximation is
averaged in the process of integration.

We have derived approximate equations (1.4) and (1.6). From these equations one can find an
approximation to $q(\rho)$ numerically. These equations yield a moment problem which has been
studied in the literature. In \ref{r6} and \ref[Section~6.2]{r4}, a quasioptimal
numerical method is given for solving moment problems with noisy data.

Note that the functions $s_l(\rho)$ differ only by
a factor independent of $\rho$ from the functions $f_l(\rho)$
defined in (1.5). This is clear from the definition of these functions
and formula (A.12) for $g_l$. Therefore one can use equations (1.4)
for the recovery of $q(r)$, and equations (1.6)
are not used below. From the definition
of $f_l(\rho)$
it follows that the set $\{f_l(\rho)\}_{0\leq l \leq L}$ of
these functions is linearly independent for any finite
positive integer $L$. Also, one can prove that the moment problem
(1.4) has at most one solution. Indeed, the corresponding
homogeneous problem (1.4), corresponding to $b_l=0$
for all $l=0,1,2,\ldots\,$, has
only the trivial solution because the set of functions
 $\{f_l(\rho)\}_{0\leq l \leq \infty}$ is complete in
$L^2(0,a)$ as follows from the result in \ref{r7} (see also
\ref{r4}). To see this, note that
the function $\{f_l(\rho)\}_{0\leq l <
\infty}$ differs only  by a factor independent of $\rho$
from the function $u_l^2(\rho)$ (the functions $u_l$ are defined in (A.1)), and the set
of these functions is the set of products of solutions
to homogeneous equation (2.2) (in Section~2) for all
$l=0,1,2,\ldots\,$. The set of these products is complete
in $L^2(0,a)$ because the set of functions
$\{u_l^2(\rho)\rho^{-2}Y_l(\alpha)Y_l(\beta)\}_{l=0,1,2,\ldots}$, $\alpha, \beta
\in S^2$, where $S^2$ is the unit sphere in $\rsp^3$,
is the set of products of solutions to the homogeneous Schr\"{o}dinger equation
which is complete in $L^2(B_a)$, as follows from the results in \ref{r4}.

Let us explain the idea of the method discussed in detail in \ref{r6}, which
is similar to the well-known Backus\,--\,Gilbert method \ref{r4}. Fix a natural
number~$L$ and look for an approximation of $q(r)$ of the form
\begin{equation}
q_L(r):=\sum^L_{l=0} b_l\nu_l(r):=\int^a_0 A_L(r,\rho)q(\rho)d\rho
\end{equation}
where the kernel $A_L$ is defined by formula (1.10) below, $b_l$ are
the known numbers given in (1.3), and $\nu_l(r)$ are not known and should
be found for any fixed $r\in [0,a]$ so that
\begin{equation}
\|q_L(r)-q(r)\| \rightarrow 0\quad \mbox{as} \quad L\rightarrow \infty.
\end{equation}
The norm in (1.9) is $L^2[0,a]$ or $C[0,a]$ norm
depending on whether $q\in L^2[0,a]$ or
$q\in C[0,a]$. Condition (1.9) holds if the sequence of the kernels
\begin{equation}
A_L(r,\rho):=\sum^L_{l=0} \nu_l(r)f_l(\rho)
\end{equation}
is a delta-sequence, that is,
\begin{equation}
A_L(r,\rho) \rightarrow \delta(r-\rho),\qq L\rightarrow +\infty
\end{equation}
where $\delta(r-\rho)$ is the delta-function.

For (1.11) to hold, we calculate $\nu_l(r)$ and $\mu_l(r)$ from the
conditions:
\begin{equation}
\int_0^a A_L(r,\rho)d \rho =1
\end{equation}
\begin{equation}
\int^a_0 |A_L(r,\rho)|^2|r-\rho|^\gamma d\rho = \min.
\end{equation}
The parameter $r\in [0,a]$ in (1.12), (1.13) is arbitrary but fixed.
Condition (1.12) is the normalization condition, condition (1.13) is the
optimality condition for the delta-sequence $A_L(r,\rho)$. It says that
$A_L(r,\rho)$ is concentrated near $r=\rho$, that is, $A_L$ is small
outside a small neighborhood of the point $\rho=r$. One can take
$\gamma=2$ in (1.13) for example.
The choice of $\gamma$ defines the degree of concentration of
$A_L(r,\rho)$ near $\rho=r$. We have taken $|A_L|^2$ in (1.13) because in
this case the minimization problem (1.12), (1.13) can be reduced to solving a linear
algebraic system of equations.

In order to calculate $q_L(r)$ by formula (1.8) one has to calculate
$\nu_l(r)$ and~$\mu_l(r)$ for different values of $r\in [0,a]$.

The problem (1.12), (1.13) is a problem of minimization of the quadratic
form (1.13) with respect to the variables $\nu_l(r)$ (which are
considered as numbers for a fixed $r$) under the linear constraint (1.12).
Such a problem can be solved, for example, by the Lagrange multipliers
method \ref[Section~6.2]{r4}.

In Section~2 we discuss numerical aspects of the proposed method for
solving the inverse scattering problem. The idea of the method is also
applicable to the inverse scattering problem with data given at a fixed
$l$ for all $k>0$, or for some  values of $k$. We specify $\psi^{(0)}_l$ in
equation (1.2) and give a method for computing the data (1.1). In Section~3
a summary of the proposed method is given. In the Appendix we have collected
all the necessary reference formulas in order to make this paper
self-contained.

In \ref{r11} numerical results obtained by the proposed approximate
inversion method are given.

\Section{Numerical aspects}
\setcounter{equation}{0}
\renewcommand{\theequation}{2.\arabic{equation}}
\vspace*{-0.5pt}   

\SubSection{Calculating $\psi_{0l}$ and the data (1.1)}
We start with a method for calculating the data (1.1) from
the original data $\{\delta_l$, $l=0,1,2,\ldots\}$ and $\{q(r)$, $r\ge a\}$.

The following integral equation is convenient for finding the data (1.1):
\begin{equation}
\psi_l(r) = \psi_{0l}(r)-\int^\infty_r \xi_l(r,\rho)q(\rho)\psi_l(\rho)d\rho, \quad r> a
\end{equation}
where the Green function $\xi_l(r,\rho)$ is defined in (A.10) and
$\psi_{0l}$ is defined by formula (2.6) below.
Equation (2.1) is a Volterra integral equation.
It can be solved stably and numerically efficiently
by iterations on the interval $r\ge a$ where
$q(r)$ is known. The solution to equation (2.1) is used in equation (2.7)
to determine $\psi_l^{(0)}$.

The function $\psi_{0l}$ is the unique solution to the equation
\begin{equation}
\psi''_{0l} + k^2\psi_{0l} - \frac{l(l+1)}{r^2} \psi_{0l} =0,\quad r>0
\end{equation}
which has the same asymptotics as $\psi_l$ as $r\rightarrow +\infty$:
\begin{equation}
\psi_l\sim e^{\i\delta_l} \sin\big(kr-\frac{l\pi}{2} + \delta_l\big),\quad r
\rightarrow\infty.
\end{equation}
This formula is derived in Appendix (formula (A.29)). Let us derive a
formula for $\psi_{0l}$. If $\psi_{0l}$ has the asymptotics (2.3), then
\begin{equation}
\psi_{0l} =c_1u_l(kr)+c_2v_l(kr)\sim c_1 \sin\big(kr-\frac{l\pi}{2}\big)-
c_2\cos\big(kr-\frac{l\pi}{2}\big),\quad r\rightarrow \infty
\end{equation}
where $u_l$ and $v_l$ are defined in (A.1) and $c_1, c_2$ are some constants.

From (2.4) and the asymptotics (2.3) for $\psi_{0l}$,  it follows that
\begin{equation}
c_1 = e^{\i\delta_l} \cos(\delta_l),\qquad c_2 = -e^{\i\delta_l}\sin(\delta_l).
\end{equation}
Thus we obtain an explicit formula for $\psi_{0l}$:
\begin{equation}
\psi_{0l}(kr) = e^{\i\delta_l}\cos(\delta_l)\,u_l(kr)-e^{\i\delta_l}\sin(\delta_l)\,
v_l(kr).
\end{equation}

\SubSection{Calculating $\psi_l^{(0)}$}
Let us find $\psi_l^{(0)}$ in equation (1.2). We start with the integral
equation (A.14) for the physical wave function $\psi_l$ and rewrite
 (A.14) as equation (1.2) with
\begin{equation}
\psi_l^{(0)} = \psi_l^{(0)}(r,k): = u_l(kr)- \int^\infty_a
g_l(r,\rho)q(\rho)\psi_l(\rho,k)\d\rho.
\end{equation}
Formula (2.7) solves the problem of finding $\psi_l^{(0)}$ in equation
(1.2). Indeed, in Section~2.1 we have shown how to calculate
$\psi_l(\rho,k)$ for $\rho\ge a$, so that the right-hand side of (2.7)
is now a known function of $r$.
This completes the description of the numerical aspects of the proposed inversion method.

\Section{Summary of the proposed method}
\setcounter{equation}{0}
\renewcommand{\theequation}{3.\arabic{equation}}
\vspace*{-0.5pt}   

Let us summarize the method:
\begin{enumerate}
\item Given the data $\{\delta_l$, $l=0,1,2,\ldots\,$; $q(r)$, $r\ge a\}$ one calculates
$\psi_{0l}$ by formula (2.6), solves equation (2.1) by iterations for $r\ge a$,
and obtains the data (1.1).
\item Given the data (1.1), one calculates $b_l$ by formulas (1.3), then chooses an 
integer $L$ and
solves the moment problem (1.4) for $0\le l \le L$. The approximate solution $q_L(r)$ is
given by formula (1.8) in which $\nu_l(r)$  are found by
solving the optimization problem (1.12), (1.13).
\end{enumerate}

In conclusion we make several remarks.

{\it Remark 1.}
This method can be formulated also for the case
when the data $b_l$  of the moment problem are known with
some random errors (see \ref[- Section~6.2]{r4} for details). This idea
can also be
tried for a numerical inversion of fixed-energy scattering data.

{\it Remark 2.}
This inverse scattering problem with fixed-energy data is highly ill-posed if $k$ is not 
very
large. Specific estimates of the conditional stability of the solution to this
problem were obtained in \ref{r5} and illustrated in \ref{r1} (see also~\ref{r4}).
Although it was proved in \ref{r7} (see also \ref{r8}--\ref{r9}) that the 3D inverse 
scattering
problem with fixed-energy scattering data has at most one solution in the class of
compactly supported potentials, but the estimates of the stability of
this solution with respect to perturbation of the scattering data
obtained in~\ref{r5} indicate that the stability is very weak: theoretically
one has to have the scattering data known with very high accuracy in
order to recover the potential with a moderate accuracy. Of course the
estimates in \ref{r5} cover the worst possible case, and in practice the
recovery may be much better than the error estimates from \ref{r5} guarantee.

In a recent paper \ref{r1} an example is given of two quite different
piecewise-constant, real-valued, compactly supported potentials which
have practically the same phase shifts at a fixed energy for all $l$ (the
phase shifts differ by a quantity of order 10$^{-5}$).

This result shows that it is necessary to have some {\it a priori}
information about the unknown potential $q(r)$ in order to be able to
recover it with some accuracy.

{\it Remark 3.}
In the Born approximation, inversion of
fixed-energy scattering data is reduced to solving an integral equation
of the form:
\begin{equation}
\int_{B_a} q(x)e^{-\i\xi\cdot x}\d x = f(\xi), \quad |\xi|\le 2k
\end{equation}
where $k>0$ is a fixed given number,
$x\in \rsp^3$, and $a>0$ is  the radius of a ball
$B_a$ outside of which $q(x)=0$. If one knows $f(\xi)$ for all $\xi\in
\rsp^3$, one  can find $q(x)$ by taking the inverse Fourier transform of
$f(\xi)$. If $f(\xi)$ is known only for $|\xi|< 2k$, one can
still recover a compactly supported $q(x)$ uniquely and analytically with
an arbitrary accuracy, if $f(\xi)$ is known exactly for $|\xi|\le 2k$
(see \ref[Section~6.1]{r4},
 where analytic inversion formulas for (3.1) are
derived).

{\it Remark 4.}
A numerical implementation of the parameter-fitting procedure of R.\,G.~Newton
requires {\it a priori} information about the potential $q$, in particular, the
knowledge of $q$ for $r\ge a$, see~\ref{r10}.

\section{Appendix}
\setcounter{equation}{0}
\renewcommand{\theequation}{A.\arabic{equation}}
\vspace*{-0.5pt}   

Equation (2.2) has linearly independent solutions
$u_l(kr)$ and $v_l(kr)$, called Riccati\,--\,Bessel functions:
\begin{equation}
u_l(r):= \sqrt{\frac{\pi r}{2}}J_{l+1/2}(r):=rj_l(r),\quad
v_l(r)=\sqrt{\frac{\pi r}{2}} N_{l+1/2}(r):= rn_l(r)
\end{equation}
where $J_\nu(r)$ and $N_\nu(r)$ are the Bessel and Neumann functions regular and
irregular at the origin respectively, $j_l$ and $n_l$ are
the spherical Bessel and Neumann functions. One has
\begin{equation}
v_l(r)\sim -\cos(r-\frac{l\pi}{2}), \quad u_l(r)\sim \sin(r-\frac{l\pi}{2}),
\q   r\rightarrow \infty
\end{equation}
\begin{equation}
v_l(r)\sim -\frac{(2l-1)!!}{r^l},\quad u_l(r)\sim\frac{r^{l+1}}{(2l+1)!!},
\q    r\rightarrow 0.
\end{equation}

The regular solution to (2.2), denoted by $\varphi_{0l}(kr)$ and defined by the condition
\begin{equation}
\varphi_{0l}(kr) \sim \frac{r^{l+1}}{(2l+1)!!},  \q r\rightarrow 0
\end{equation}
is given by the formula
\begin{equation}
 \varphi_{0l}(kr) = \frac{u_l(kr)}{k^{l+1}}
\end{equation}
which follows from (A.3) and (A.4).

The Jost solution $f_{0l}$ to (2.2) defined by the condition
\begin{equation}
f_{0l}(kr) \sim e^{\i kr}, \q r\rightarrow \infty
\end{equation}
is given by the formula
\begin{equation}
f_{0l}(kr) = \e^{\i(l+1)\pi/2} u_l(kr)-\e^{\i l\pi/2}
v_l(kr)=\i\e^{\i l\pi/2}(u_l +\i v_l)
\end{equation}
as follows from (A.2).

The Wronskian is
\begin{equation}
F_{0l}(k): =W[f_{0l},\varphi_{0l}] = f_{0l}(kr)\frac{\ddd}{\ddd r}
 \varphi_{0l}(kr)-\varphi_{0l}(kr)\frac{\ddd}{\ddd r}f_{0l}(kr) =
 \frac{\e^{\i l\pi/2}}{k^l}
\end{equation}
as follows from (A.3), (A.4), (A.7), and the fact that this Wronskian
does not depend on $r$, so it can be calculated at $r=0$.

In equation (2.1) we use the Green function $\xi_l(r,\rho)$ satisfying the equation
\begin{equation}
\frac{\ddd^2\xi_l(r,\rho)}{\ddd r^2}+k^2\xi_l(r,\rho)-\frac{l(l+1)}{r^2}
\xi_l(r,\rho) = -\delta(r-\rho)
\end{equation}
where $\delta(r-\rho)$ is the delta-function, and vanishing for $\rho <r$:
\begin{equation}
\xi_l(r,\rho) = \left\{\begin{array}{cl}
\e^{-\i l\pi/2}k^l[-\varphi_{0l}(k\rho)f_{0l}(kr)+f_{0l}(k\rho)
  \varphi_{0l}(kr)],\q & \rho\ge r\\
  0, & \rho <r.
 \end{array}\right.
\end{equation}
Indeed, $\xi_l$ solves (A.9) for $r\not=\rho$ since, as a function of
$r$, it is a linear combination of solutions to the homogeneous equation
(A.9). Moreover,
\begin{equation}
\predel{\frac{\ddd\xi_l(r,\rho)}{\ddd r}}^{r=\rho +0}_{r=\rho -0}
=\e^{-\i l\pi/2}k^l
\predel{\Big[\varphi_{0l}(k\rho)\frac{\ddd f_{0l}(kr)}{\ddd r}
-f_{0l}(k\rho)\frac{\ddd\varphi_{0l}(kr)}{\ddd r}\Big]}_{r=\rho}=-1
\end{equation}
by (A.8). Thus (A.9) follows.

The Green function
\begin{equation}
g_l(r,\rho):=\left\{\begin{array}{c@{\q}l}
F^{-1}_{0l}(k)\varphi_{0l}(k\rho)f_{0l}(kr),& r\ge\rho\\
F^{-1}_{0l}(k)\varphi_{0l}(kr)f_{0l}(k\rho),& r<\rho
\end{array} \right.
\end{equation}
(where $F_{0l}(k)$ is defined in (A.8), $F_{0l}=\e^{\i l\pi/2}k^{-l}$) solves (A.9)
and satisfies the radiation condition at infinity:
\begin{equation}
\frac{\partial g_l}{\partial r} -\i kg_l \rightarrow 0 \q\mbox{as}
\quad r\rightarrow +\infty.
\end{equation}
Indeed, (A.13) follows from (A.12) and (A.6), and to verify that $g_l$
solves equation (A.9) one argues in the same way as was done
after formula (A.10).

The integral equation for the physical wave function $\psi_l(r,k)$ is:
\begin{equation}
\psi_l(r,k) = u_l(kr)-\int^\infty_0 g_l(r,\rho)q(\rho)\psi_l(\rho,k)\d\rho.
\end{equation}
This equation follows from the standard three-dimensional equation:
\begin{equation}
\psi(x,\alpha,k) = \psi_0(x,\alpha,k)-
\int_{\rsp^3}\frac{\e^{\i k|x-y|}}{4\pi|x-y|}
q(\rho)\psi (y, \alpha, k) \d y, \q \rho=|y|,\ r=|x|
\end{equation}
where  $\psi_0(x,\alpha, k): =\e^{\i k\alpha\cdot x}$,  and $\alpha\in S^2$ is a given
unit vector.

If one writes
\begin{equation}
\e^{\i k\alpha\cdot x} = \sum_{l\ge 0}
\frac{4\pi}{k}\i^l\frac{u_l(kr)}{r}Y_l(x^0)\overline{Y_l(\alpha)},\q x^0:=\frac{x}
{|x|}
\end{equation}
\begin{equation}
\psi = \sum_{l\ge 0}\frac{4\pi}{k}\i^l\frac{\psi_l(r,k)}{r} Y_l(x^0)\overline{Y_l(\alpha)}
\end{equation}
\begin{equation}
\frac{\e^{\i k|x-y|}}{4\pi|x-y|} = \sum_{l\ge 0}
\frac{g_l(r,\rho)}{r\rho} Y_l(x^0)\overline{Y_l(y^0)}
\end{equation}
where $Y_l:=Y_{lm}$, $-l\leq m \leq l$, are the orthonormal spherical harmonics, and inserts
(A.16)--(A.18) into (A.15), then one gets (A.14).
In (A.16)--(A.18) the summation with respect to $l$, $l=0,1,2,3,\ldots\,$,
includes also the summation with respect to $m$, $-l\leq m \leq l$.

Let $r\rightarrow +\infty$ in (A.14). Then (A.2), (A.5), (A.6), (A.8), and (A.12) imply
\begin{equation}
\psi_l\sim \frac{\e^{\i\pi(l+1)/2}}{2}[\e^{-\i kr}-\e^{\i\pi l}\e^{\i kr}S_l],
\q r\rightarrow +\infty
\end{equation}
where
\begin{equation}
S_l=1+\frac{2}{\i k}\int^\infty_0 u_l(k\rho)q(\rho)\psi_l(\rho,k)\d\rho.
\end{equation}

The scattering amplitude $A(\alpha',\alpha,k)$ defined by the formula
\begin{equation}
\psi=\exp(\i k\alpha\cdot x)+A(\alpha',\alpha,k)
\frac{\e^{\i kr}}{r} +o\big(\frac{1}{r}\big),
 \q r:=|x|\rightarrow\infty,
 \quad \frac{x}{|x|}=\alpha'
\end{equation}
can be written as
\begin{equation}
A(\alpha ',\alpha,k)=-\frac{1}{4\pi}\int_{\rsp^3} \e^{-\i k\alpha'\cdot y}
q(\rho)\psi(y,\alpha,k) \d y.
\end{equation}
If $q=q(\rho)$, $\rho :=|y|$, then
\begin{equation}
A(\alpha',\alpha ,k)= \sum^\infty_{l=0}A_l(k)Y_l(\alpha')\overline{Y_l(\alpha)}
\end{equation}
where
\begin{equation}
A_l(k) =-\frac{4\pi}{k^2} \int^\infty_0 u_l(k\rho)q(\rho)\psi_l(\rho,k)\d\rho
\end{equation}
as one gets after substituting the complex conjugate of (A.16), (A.17), and~(A.23)
into (A.22). From (A.20) and (A.24) one gets
\begin{equation}
S_l = 1-\frac{k}{2\pi \i}A_l
\end{equation}
which is the fundamental relation between the $S$-matrix and the
scattering amplitude:
\begin{equation}
S = I-\frac{k}{2\pi \i} A.
\end{equation}
Here $S$, $I$, and $A$  are operators in $L^2(S^2)$, $I$ is the
identity operator. Since the operator $S$ is unitary in $L^2(S^2)$ and
$S_l$ are the eigenvalues of $S$ in the eigenbasis of spherical
harmonics in the case of spherically symmetric potentials, one has
$|S_l|=1$, so, for some real number $\delta_l$ one writes
\begin{equation}
S_l = \e^{2\i\delta_l}.
\end{equation}
The numbers $\delta_l$ normalized by the condition
\begin{equation}
\delta_l(k) \rightarrow 0 \quad\mbox{as} \quad k\rightarrow\infty
\end{equation}
are called the phase shifts. They are in one-to-one correspondence with
the numbers $S_l$ if (A.28) is assumed. We do assume (A.28).

Formula (A.19) can be rewritten as
\begin{equation}
\psi_l \sim \e^{\i\delta_l}\frac{\e^{\i(kr-l\pi/2+\delta_l)} -
\e^{-\i(kr-l\pi/2+\delta_l)}}{2\i}
= \e^{\i\delta_l} \sin\big(kr-\frac{l\pi}{2} +\delta_l\big), \q r\rightarrow \infty.
\end{equation}
Formula (A.29) was used in Section~2 (see formula (2.3)).

Note that (A.25) and (A.27) imply
\begin{equation}
 A_l=\frac {4\pi}{k} \e^{\i\delta_l} \sin(\delta_l).
\end{equation}
This formula and (A.23) show that if $q(x)=q(|x|)$,
then the information which is given by the knowledge
of the fixed-energy phase shifts $\{\delta_l$, $l=0,1,2,\ldots\,\}$
is equivalent to the information which is given by
the knowledge of the fixed-energy scattering amplitude
$A(\alpha', \alpha)$ where $\alpha'$ and $\alpha$ run through
the whole of the unit sphere $S^2$. By Ramm's uniqueness theorem
\ref{r7}, it follows that if the fixed-energy phase shifts
corresponding to a compactly supported potential
$q(r)$ are known for $l=0,1,2,3,\ldots\,$, then the potential $q(r)$ is
uniquely defined, and therefore the phase shifts $\delta_l$ are
uniquely defined for all $l$ on the complex
plane by their values for $l$ running through the integers only: $l=0,1,2,\ldots\,$.

{\bf Acknowledgement}

This paper was written while A.\,G.~Ramm was visiting the Institute of Theoretical
Physics at the University of Giessen.
AGR thanks DAAD and this University for hospitality.

\end{document}